\documentclass[a4paper,12pt, epsfig]{article} \def\letter{0}\def\pr{0}
\usepackage{epsfig}
\usepackage{epstopdf}
\usepackage{graphicx}
\usepackage{ifthen}

\usepackage{amsmath}
\usepackage[psamsfonts]{amssymb}
\usepackage{euscript}

\usepackage{latexsym}
\usepackage[arrow,matrix,curve]{xy}

\newskip\humongous \humongous=0pt plus 1000pt minus 1000pt

\newif\ifdtup

\def\,{\hspace{-.1cm}}
\def\hsp{,\hspace{.7cm}}

\def\fc#1#2 {\frac{n}{q}#1\frac{n}{q}#2}

\def\kt{\kappa}
\def\kt{\mathfrak{K}}
\def\ks{|\kt\rangle}
\def\bdk{B^\dag_\kt}

\newcommand{\vac}{\ensuremath{|0\rangle}}

\renewcommand{\sin}{\textrm{sin}}

\renewcommand{\tanh}{\textrm{tanh}}
\newcommand{\sech}{\textrm{sech}}
\newcommand{\csch}{\textrm{csch}}

\def\cc{\mathcal{C}}

\renewcommand{\theequation}{\arabic{section}.\arabic{equation}}
\renewcommand{\(}{\begin{equation}}
\renewcommand{\)}{end{equation} \vspace{-.05in}\linebreak}

\newcounter{saveeqn}
\newcounter{savealpheqn}

\newcommand{\alpheqn}{\setcounter{saveeqn}{\value{equation}}%
  \stepcounter{saveeqn}\setcounter{equation}{0}%
  \renewcommand{\theequation}{\mbox{\arabic{section}.\arabic{saveeqn}
\alph{equation}}}
  \renewcommand{\)}{\end{equation}}}
\def\part#1{\frac{\partial}{\partial{#1}}}%
\def\group#1{\refstepcounter{equation}\setcounter{saveeqn}
 {\value{equation}}%
  \label{#1}\setcounter{equation}{0}%
\renewcommand{\theequation}{\mbox{\arabic{section}.\arabic{saveeqn}
\alph{equation}}}
  \renewcommand{\)}{\end{equation}}}
\newcommand{\reseteqn}{\setcounter{equation}{\value{saveeqn}}%
  \renewcommand{\theequation}{\arabic{section}.\arabic{equation}}%
  \renewcommand{\)}{\end{equation}}}

\newcommand{\aalpheqn}{\setcounter{saveeqn}{\value{equation}}%
  \stepcounter{saveeqn}\setcounter{equation}{0}%
  \renewcommand{\theequation}{\mbox{
        \Alph{subsection}.\arabic{saveeqn}\alph{equation}}}
   \renewcommand{\)}{\end{equation}}}
\newcommand{\areseteqn}{\setcounter{equation}{\value{saveeqn}}%
  \renewcommand{\theequation}{\Alph{subsection}.\arabic{equation}}%
  \renewcommand{\)}{\end{equation}}}

\renewcommand{\thefootnote}{\alph{footnote}}
\renewcommand{\(}{\begin{equation}}
\renewcommand{\)}{\end{equation}}
\newcommand{\ba}{\begin{eqnarray}}
\newcommand{\ea}{\end{eqnarray}}
\renewcommand{\a}{\alpha}
\renewcommand{\b}{\beta}

\newcommand{\cbp}{\mathop{\vtop{\ialign{##\crcr
   $\hfil\displaystyle{}\hfil$\crcr\noalign{\kern-13pt\nointerlineskip}
   \BIG{)}\hskip0pt\crcr\noalign{\kern3pt}}}}}
\newcommand{\pa}{\mathop{\vtop{\ialign{##\crcr

$\hfil\displaystyle{\oplus}\hfil$\crcr\noalign{\kern+1pt\nointerlineskip
}
   \hspace{.08in}$^{\alpha=0}$\hskip6pt\crcr\noalign{\kern3pt}}}}}
\renewcommand{\hsp}{,\hspace{.3in}}
\newcommand{\p}{^\prime}
\newcommand{\pp}{^{\prime\prime}}

\newcommand{\Z}{\ensuremath{\mathbb Z}}

\def\rd{\rho^{\rm{}}}
\def\rv{\rho^{\rm{(vac)}}}
\def\rdiv{\rho^{\rm{(div)}}}
\def\rfin{\rho^{\rm{(fin)}}}
\def\rdif{\rho^{\rm{(dif)}}}
\def\qfin{Q_2^{\rm{(33fin)}}}

\catcode`\@=11
\def\vereq#1#2{\lower3pt\vbox{\baselineskip1.5pt \lineskip1.5pt
\ialign{$\m@th#1\hfill##\hfil$\crcr#2\crcr\sim\crcr}}}
\catcode`\@=12

\renewcommand{\(}{\begin{equation}}
\renewcommand{\)}{\end{equation}}

\def\pin#1{\int \frac{d#1}{2\pi}}
\def\pink#1{\int \frac{d^{#1}k}{(2\pi)^{#1}}}

\def\pinkp#1{\int \frac{d^{#1}k\p}{(2\pi)^{#1}}}

\def\dpin#1{\int^+ \frac{d#1}{2\pi}}
\def\dpink#1{\int^+ \frac{d^{#1}k}{(2\pi)^{#1}}}

\def\Bd#1{B^\dag_{k_{#1}}}

\def\df{\mathcal{D}_f}
\def\B#1{B^\dag_{k_{#1}}}

\def\I{\mathcal{I}}
\def\vl#1#2#3{\vector(#1,#2){#3}\line(#1,#2){#3}}
\def\wk#1{\omega_{k_{#1}}}
\def\wkp#1{\omega_{k^{'}_{#1}}}

\newcommand{\beas}{\begin{eqnarray*}}
\newcommand{\eeas}{\end{eqnarray*}}

\newcommand{\bquo}{\begin{quote}}
\newcommand{\enqu}{\end{quote}}

\renewcommand{\Z}{{\mathbb Z}}

\def\ch{{\mathcal{H}}}

\def\ok#1{\omega_{k_{#1}}}
\def\okp#1{\omega_{k\p_{#1}}}
\def\V#1{V^{(#1)}[\sqrt{\lambda}f(x)]}
\def\ck{\csch\left(\frac{\pi k}{2\b}\right)}
\def\cks{\csch^2\left(\frac{\pi k}{2\b}\right)}
\def\sk{\sech\left(\frac{\pi k}{2\b}\right)}
\def\sks{\sech^2\left(\frac{\pi k}{2\b}\right)}
\def\kkk{\sum_{i}^3k_i}
\def\csk{\csch\left(\frac{\pi\kkk}{2\b}\right)}
\def\csks{\csch^2\left(\frac{\pi\kkk}{2\b}\right)}

\def\hp{{\hat{\Phi}}}

\newcommand{\beq}{\begin{equation}}
\newcommand{\eeq}{\end{equation}}
\newcommand{\bea}{\begin{eqnarray}}
\newcommand{\eea}{\end{eqnarray}}

\newskip\humongous \humongous=0pt plus 1000pt minus 1000pt

\newif\ifdtup

\catcode`\@=11

\ifthenelse{\equal{\letter}{0}}{ 

\@addtoreset{equation}{section}
\def\theequation{\arabic{section}.\arabic{equation}}

\def\@normalsize{\@setsize\normalsize{15pt}\xiipt\@xiipt
\abovedisplayskip 14pt plus3pt minus3pt%
\belowdisplayskip \abovedisplayskip
\abovedisplayshortskip \z@ plus3pt%
\belowdisplayshortskip 7pt plus3.5pt minus0pt}

\def\small{\@setsize\small{13.6pt}\xipt\@xipt
\abovedisplayskip 13pt plus3pt minus3pt%
\belowdisplayskip \abovedisplayskip
\abovedisplayshortskip \z@ plus3pt%
\belowdisplayshortskip 7pt plus3.5pt minus0pt
\def\@listi{\parsep 4.5pt plus 2pt minus 1pt
      \itemsep \parsep
      \topsep 9pt plus 3pt minus 3pt}}

\relax

\def\section{\@startsection{section}{1}{\z@}{3.5ex plus 1ex minus  .2ex}{2.3ex plus .2ex}{\large\bf}}

\def\thesection{\arabic{section}}
\def\thesubsection{\arabic{section}.\arabic{subsection}}

\def\appendix{\setcounter{section}{0}
 \def\thesection{Appendix \Alph{section}}
 \def\thesubsection{\Alph{section}.\arabic{subsection}}
 \def\theequation{\Alph{section}.\arabic{equation}}}
\renewcommand{\theequation}{\arabic{section}.\arabic{equation}}

}{
\renewcommand{\theequation}{\arabic{equation}}

} 

\begin{document}
\def\thefootnote{\fnsymbol{footnote}}
\def\thetitle{The $\phi^4$ Kink Mass at Two Loops}
\def\autone{Jarah Evslin}
\def\affa{Institute of Modern Physics, NanChangLu 509, Lanzhou 730000, China}
\def\affb{University of the Chinese Academy of Sciences, YuQuanLu 19A, Beijing 100049, China}

\title{Titolo}

\ifthenelse{\equal{\pr}{1}}{
\title{\thetitle}
\author{\autone}
\affiliation {\affa}
\affiliation {\affb}

}{

\begin{center}
{\large {\bf \thetitle}}

\bigskip

\bigskip


{\large \noindent  \autone{${}^{1,2}$}}


\vskip.7cm

1) \affa\\
2) \affb\\

\end{center}

}

\begin{abstract}
\noindent
The two-loop correction to the mass of the $\phi^4$ kink is $0.0126\lambda/m$ in terms of the coupling $\lambda$ and the meson mass $m$ evaluated at the minimum of the potential.  This is calculated using a recently proposed alternative to collective coordinates.  Both the kink energy and the vacuum energy are IR divergent at this order.  To cancel the divergence, the two energy densities are subtracted before integrating over space, or equivalently a finite counterterm is added to the Hamiltonian density to cancel the vacuum energy density.  All spatial integrals are performed analytically.  However in the last step of our calculation, integrals over virtual momenta are performed numerically.



\noindent

\end{abstract}

%
\setcounter{footnote}{0}
\renewcommand{\thefootnote}{\arabic{footnote}}

\ifthenelse{\equal{\pr}{1}}{
\maketitle
}{}

\section{Introduction}

The $\phi^4$ double well model in 1+1 dimensions has two vacua.  In each, a real scalar field has a mass of $m=2\b$.  A classical kink solution of mass
\beq
Q_0=\frac{8\b^3}{3\lambda}
\eeq
 interpolates between the two vacua.  The one-loop mass, defined as the difference between the kink state energy and the vacuum energy, was calculated in \cite{dhn2} to be
\beq
Q_1=\left(-\frac{3}{\pi}+\frac{1}{2\sqrt{3}}\right)\b. \label{q1}
\eeq
The kink state energy and vacuum energy were calculated using the kink Hamiltonian and vacuum Hamiltonian respectively.  In principle the theory is defined by the vacuum Hamiltonian, but the use of the kink Hamiltonian to calculate the kink state energy is justified by the fact that the two Hamiltonians have the same spectrum.  

The problem is that the two Hamiltonians need to be regularized, and regularization changes their spectra.  After regularization, the energies depend on the regulators and, as explained in \cite{rebhan}, even for the same value of the regulator, the kink state energy calculated using the kink Hamiltonian may no longer be equal to that calculated using the defining vacuum Hamiltonian.  It is of course the eigenvalue of the defining Hamiltonian which gives the mass, and so if these two eigenvalues are not equal, then the calculation that uses the kink Hamiltonian will obtain the wrong answer.  In Ref.~\cite{rebhan} it was shown that for some regulators this mismatch persists even when the regulator is taken to infinity.  Worse yet, while there have been many proposed principles in the literature \cite{lit,local}, there is no established criterion for determining which regulators lead to this mismatch\footnote{Of course one can obtain nonperturbative results on the lattice or using Hamiltonian truncation or Borel resummation, without recourse to a kink Hamiltonian, but that will not be our approach here.  We will see in Sec.~\ref{compsez} that our method yields superior precision at weak coupling.}.  

This ambiguity can be avoided when the kink mass is fixed by another principle, such as integrability or supersymmetry, if the regularization preserves this property.  However the $\phi^4$ theory has neither property.  The ambiguity can also be avoided at one loop, where the problem reduces to finding the density of states of a free theory \cite{gj}, and so it was eventually shown that the one-loop result (\ref{q1}) is nonetheless correct.

What about two loops?  The goal of the present paper will be to calculate the two-loop mass of the $\phi^4$ kink.  The following trick was introduced in Ref.~\cite{mekink} in the context of this model and generalized to other models in \cite{memassa}.   One observes that the kink Hamiltonian $H\p$ can be defined via a similarity transformation of the original Hamiltian $H$
\beq
H\p=\df^{-1} H\df \label{sim}
\eeq
where $\df$ is a unitary operator.  Two similar operators necessarily have the same spectrum.  The key innovation in this procedure is that one first regularizes $H$ and then defines the regularized $H\p$ via (\ref{sim}).  Now there is a single regulator, and the spectrum of $H\p$ is equal to that of the defining $H$ at each value of the regulator, so that the energy calculated using $H\p$ agrees with the correct energy, which would be obtained using $H$.  Thus the problem is solved.  In Ref.~\cite{me2loop} this method was used to find the two-loop energy of a general kink in its ground state.

Unfortunately, in the case of the $\phi^4$ theory, and any theory whose potential has a nonvanishing third derivative at its minima, this energy suffers from an infrared (IR) divergence.  This divergence is to be expected, because the vacuum at this order has a constant energy density of \cite{slava,serone,hui}
\beq
\rv(x)=\left(\frac{1}{24}-\frac{\psi^{(1)} (1/3)}{16\pi^2}
\right)\lambda
\sim -0.0222644\lambda
\eeq
and so an infinite energy.  Here $\psi^{(1)}(z)=\partial_z^2 {\rm{ln}}\left(\Gamma(z)\right)$ is the polygamma function.  The mass, which is the difference between these two infinite quantities, is expected to be finite.  

How can these two infinite quantities be reliably subtracted?   Usually one removes IR divergences by calculating an experimentally accessible quantity.  One might be tempted to ask how much energy one needs to create a kink.  Unfortunately this is impossible in the quantum theory \cite{mandal} as the kink changes the boundary conditions, creating a lone soliton would require an infinite action.   One could also remove IR divergences by putting a system in a box, but it is known that boundary conditions can lead to mass shifts which persist even in the limit that the box size is taken to infinity.   

One is always free to set the exact vacuum energy to zero by including a counterterm  in the defining Hamiltonian density which exactly cancels the vacuum energy density \cite{colemanqft}.  As this counterterm is a $c$-number, Eq.~(\ref{sim}) implies that it will also appear in the kink Hamiltonian density.  Our procedure is equivalent to modifying the calculation of \cite{me2loop} by including this counterterm, defined by the condition that the vacuum energy vanishes.

While equivalent, we find that it is more convenient to subtract the vacuum energy density from the kink energy density rather than to subtract it directly from the kink Hamiltonian density.  This requires an {\it{ad hoc}} convention for the definition of the kink energy density, but after integration over space the result is independent of this convention.   The complication is that the kink is not an eigenstate of the Hamiltonian density operator, nor is the vacuum.  One might try to define the energy density as the expectation value of the Hamiltonian density operator, but alas the Hamiltonian eigenstates have constant momenta and so are not normalizable.   This could in principle be resolved by introducing wave packets, but a finite width wave packet would increase the energy and in the limit that the width goes to infinity, the divergence would return.  In principle one could nonetheless subtract the vacuum energy inside of the support of the wave packet as the wave packet size goes to infinity, but this procedure would be ambiguous as a finite shift in the definition of the support of the wave packet would lead to a finite shift in the limit, and so in the calculated mass.

We instead adopt the following definition.   If $\vac$ is the kink ground state after the similarity transform (\ref{sim}), then 
\beq
H\p\vac=Q\vac. \label{seq}
\eeq
If $H\p$ is the integral of a Hamiltonian density $\ch\p(x)$, then we can expand $\ch\p(x)\vac$ in a basis of the Hilbert space, with $\vac$ an element of the basis.  Integrating over $x$ this must reduce to (\ref{seq}) and so the coefficients of all basis elements except for $\vac$ will integrate to zero while the coefficient of $\vac$ integrates to $Q$.  Since the coefficient of $\vac$ in the expansion of $\ch\p(x)\vac$ integrates to $Q$, we define it to be the energy density $\rho(x)$.  While this definition of $\rho(x)$ may depend on the choice of basis, as $\rho(x)$ may mix with the coefficients of some other basis elements under a basis transformation, this ambiguity should disappear after the integration over $x$ as all other coefficients integrate to zero.

In the case of the Sine-Gordon model, at this order the correction to the vacuum energy vanishes and so this issue does not arise.  Also the two-loop kink mass is known using integrability \cite{dhn2loops,luther}.  However, one could have shifted the defining Hamiltonian density by a constant, leading to a finite vacuum energy density and an IR divergence in the kink energy.  In that case, our prescription would be equivalent to simply shifting the Hamiltonian back by this constant, which of course is the correct way to remove this spurious divergence.  

We begin in Sec.~\ref{revsez} with a review of the calculation of the two-loop kink ground state energy for a general kink.  In Sec.~\ref{finsez} we restrict our attention to the $\phi^4$ double well and evaluate the finite terms, reducing the problem to a finite-dimensional integral of elementary functions.  In Sec.~\ref{irsez} we treat the IR divergent terms, subtracting the corresponding vacuum energy as described above.  As a result, the entire energy is computed in the form of a three-dimensional integral over elementary functions.  In Sec.~\ref{numsez} this integral is evaluated numerically.  Finally, the result is compared with the literature in Sec.~\ref{compsez}.  Some of the most important notation is summarized in Table~\ref{notab}.

\section{Review} \label{revsez}

\begin{table}
\begin{tabular}{|l|l|}
\hline
Operator&Description\\
\hline
$\phi(x),\ \pi(x)$&The real scalar field and its conjugate momentum\\
$b^\dag_k,\ b_k,\ B^\dag_k,\ B_k$&Creation and annihilation operators in normal mode basis\\
$\phi_0$&Zero mode of $\phi(x)$ in normal mode basis\\
$::_a,\ ::_b$&Normal ordering with respect to $a$ or $b$ operators respectively\\
\hline
\hline
Hamiltonian&Description\\
\hline
$H$&The original Hamiltonian\\
$H^\prime$&$H$ with $\phi(x)$ shifted by kink solution $f(x)$\\
$H_n$&The $\phi^n$ term in $H^\prime$\\
\hline
Symbol&Description\\
\hline
$\beta$&Half of the scalar mass\\
$\lambda$&Coupling constant\\
$f(x)$&The classical kink solution\\
$\df$&Operator that translates $\phi(x)$ by the classical kink solution\\
$g_B(x)$&The kink linearized translation mode\\
$g_k(x)$&Continuum normal mode or shape mode\\
$g_S(x)$&Shape mode\\
$\gamma_i^{mn}$&Coefficient of $\phi_0^m B^{\dag n}\vac_0$ in order $i$ ground state $\vac_i$\\
$V_{ijk}$&Derivative of the potential contracted with various functions\\
$\cc_{ijk}$&Coefficient which appears in all continuous $V_{ijk}$\\
$\a_{ijk}$&Coefficient which appears in contribution of $V_{ijk}$ to energy\\
$\sigma_{ijk}(x)$&Quantity which integrates to $V_{ijk}$\\
$\Phi_{ijk}^{IJ}(x)$&Matrix characterizing $x$-dependence of $\sigma_{ijk}$\\
$\Delta_{ij}$&Integral of $g_i(x)g_j\p(x)$\\
$\rd(x)$&Energy density arising from three virtual continuum modes\\
$\rv(x)$&Vacuum energy density\\
$\I(x)$&Contraction factor from Wick's theorem\\
$k_i$&The analog of momentum for normal modes\\
$\omega_k$&The frequency corresponding to $k$\\
$Q_n$&$n$-loop correction to kink energy\\
\hline
State&Description\\
\hline
$|K\rangle, \ |\Omega\rangle$&Kink and vacuum sector ground states\\
$\vac$&Translation of $|K\rangle$ by $\df^{-1}$\\
$\vac_i$&Translation of $|K\rangle$ by $\df^{-1}$  at order $i$ \\
\hline

\end{tabular}
\caption{Summary of Notation}\label{notab}
\end{table}

Consider a (1+1)-dimensional theory of a real scalar field $\phi$ and its conjugate momentum $\pi$, defined by the Hamiltonian
\bea
H&=&\int dx \ch(x) \label{hd}\\
\ch(x)&=&\frac{1}{2}:\pi(x)\pi(x):_a+\frac{1}{2}:\partial_x\phi(x)\partial_x\phi(x):_a+\frac{1}{\lambda}:V[\sqrt{\lambda}\phi(x)]:_a\nonumber
\eea
where $::_a$ is normal-ordering of the usual operators that create and annihilate plane waves and we always work in the Schrodinger picture.  The classical equations of motion admit the kink solution
\beq
\phi(x,t)=f(x). \label{fd}
\eeq

To perturbatively treat oscillations about the kink, one would like to expand about the kink solution.  This may be done using the passive transformation of the fields $\phi\rightarrow \phi\p=\phi-f$ or else the active transformation of the functionals acting on the fields
\beq
F[\phi]\rightarrow F\p[\phi]=F[\phi\p]. \label{fp}
\eeq
We opt for the second approach, realized as follows.  Defining the displacement operator $\df$
\beq
\df={\rm{exp}}\left(-i\int dx f(x)\pi(x)\right) \label{df}
\eeq
we define the kink Hamiltonian and kink momentum as
\beq
H\p=\df^\dag H\df\hsp P\p=\df^\dag P\df.
 \label{hpd}
\eeq
The theory is rendered UV finite by normal ordering, and so (\ref{fp}) and (\ref{hpd}) are equivalent.  However, our procedure may be implemented with a general regulator and in that case (\ref{hpd}) should be taken as our definition of kink sector operators, as the similarity transform guarantees that kink sector operators will have the same spectra as the original operators.

A quick calculation \cite{mekink} shows
\bea
H\p&=&\df^\dag H\df=Q_0+\sum_{n=2}^{\infty}H_n\hsp
H_{n(>2)}=\frac{1}{n!}\int dx \V{n}:\phi^n(x):_a\\
H_2&=&\frac{1}{2}\int dx\left[:\pi^2(x):_a+:\left(\partial_x\phi(x)\right)^2:_a+V^{\prime\prime}[gf(x)]:\phi^2(x):_a\right.].\nonumber
\eea

Let $|K\rangle$ be the kink ground state, and $Q$ its energy
\beq
H|K\rangle=Q|K\rangle\hsp P|K\rangle=0.
\eeq
Then we may define
\beq
\vac=\df^\dag |K\rangle
\eeq
which is an eigenstate of the kink Hamiltonian and momentum
\beq
H\p\vac=Q\vac\hsp P\p\vac=0.
\eeq
Define a semiclassical expansion of this state and its eigenvalue
\beq
\vac=\sum_{i=0}\vac_i\hsp
Q=\sum_{j=0}Q_j
\eeq
where $\vac_i$ is the $i$th order ground state and $Q_j$ is the $j$-loop correction to its mass.  At $j$ loops the state is determined up to $i=2j-2$.

In this note we will be interested in the two-loop correction to the energy of the kink ground state \cite{me2loop}
\bea
Q_2&=&\sum_{j=1}^5 Q_2^{(j)}\hsp
Q_2^{(1)}=\frac{V_{\I \I}}{8}\hsp
Q_2^{(2)}=-\frac{1}{8}\dpin{k}\frac{\left|V_{\I  k}\right|^2}{\ok{}^2}\label{q2}\\
Q_2^{(3)}&=&-\frac{1}{48}\dpink{3} \frac{\left|V_{k_1k_2k_3}\right|^2}{\omega_{k_1}\omega_{k_2}\omega_{k_3}\left(\omega_{k_1}+\omega_{k_2}+\omega_{k_3}\right)}\nonumber\\
Q_2^{(4)}&=&\frac{1}{16Q_0}\dpink{2}\frac{\left|\left(\omega_{k_1}-\omega_{k_2}\right)\Delta_{k_1k_2}\right|^2}{\omega_{k_1}\omega_{k_2}}=\frac{1}{16}\dpink{2}\frac{\left|V_{Bk_1k_2}\right|^2}{\omega_{k_1}\omega_{k_2}\left(\ok1+\ok2\right)^2} \nonumber\\
Q_2^{(5)}&=& -\frac{1}{8Q_0^2}\int dx
\left|f^{\prime\prime}(x)\right|^2=-\frac{1}{8Q_0}\dpin{k}\left|\Delta_{k B}\right|^2= -\frac{1}{8}\dpin{k}\frac{\left|V_{BBk}\right|^2}{\ok{}^4}.
\nonumber
\eea
Here we have introduced the matrix
\beq
\Delta_{ij}=\int dx g_i(x) g\p_j(x)
\eeq
and the symbol
\beq
V_{\I\stackrel{m}{\cdots}\I,\alpha_1\cdots\alpha_n}=\int dx V^{(2m+n)}[\sqrt{\lambda}f(x)]\I^m(x) g_{\alpha_1}(x)\cdots g_{\alpha_n(x)} \label{vf}
\eeq
where $V^{(n)}$ is the $n$th derivative of $\lambda^{n/2-1}V$ with respect to its argument and $g_k(x)$ are continuous and discrete normal modes of frequency $\omega_k$, which solve the linearized equations of motion for $H\p$.  In particular $g_B(x)$ is the zero mode with $\omega_B=0$.  The normalization and phases of the normal modes are chosen so that
\bea
\int dx g_{k_1} (x) g^*_{k_2}(x)&=&2\pi \delta(k_1-k_2)\hsp 
\int dx |g_S(x)|^2=
\int dx |g_{B}(x)|^2=1\nonumber\\
g_k(-x)&=&g_k^*(x)=g_{-k}(x)\hsp g_S(-x)=g_S^*(x)
\eea
where $g_S(x)$ is any discrete normal mode and $k_i$ are real.

We have also introduced the notation that $\int^+ dk/2\pi$ is an integral over all real values of $k$ corresponding to the continuous normal modes as well as a discrete sum over the imaginary values of $k$ corresponding to discrete normal modes, like the shape mode of the $\phi^4$ kink.  The zero mode is not included in this sum.   

\begin{figure}
\setlength{\unitlength}{.7cm}
\begin{picture}(6,4)(-7,-1)
\put(-2,1){\circle{2}}
\put(-4,1){\circle{2}}
\put(-3.01,1){\circle*{.1}}

\put(6,1){\circle{2}}
\put(5,1){\vl{-1}0 {.5}}
\put(3,1){\circle{2}}
\put(4.6,1.3){\makebox(0,0){{$k\p$}}}

\qbezier(10,1)(12,3)(14,1)
\put(14,1){\vl{-1}0 2}
\put(12,2){\vector(-1,0) 0}
\put(12,0){\vector(-1,0) 0}
\qbezier(10,1)(12,-1)(14,1)
\put(12,2.4){\makebox(0,0){{$k\p_1$}}}
\put(12,1.4){\makebox(0,0){{$k\p_2$}}}
\put(12,0.4){\makebox(0,0){{$k\p_3$}}}




\end{picture}
\caption{$Q_2$ in pictures, as described in Ref.~\cite{normal}.  Each vertex represents an interaction in $H\p$, with operator ordering running to the left and a factor of $\I$ for each loop at a single vertex.  The three diagrams correspond to $Q_2^{(1)}$, $Q_2^{(2)}$ and $Q_2^{(3)}$ respectively while $Q_2^{(4)}$ and $Q_2^{(5)}$ arise from replacing a normal mode with one or two zero modes in the last diagram.}
\label{q2fig}
\end{figure}
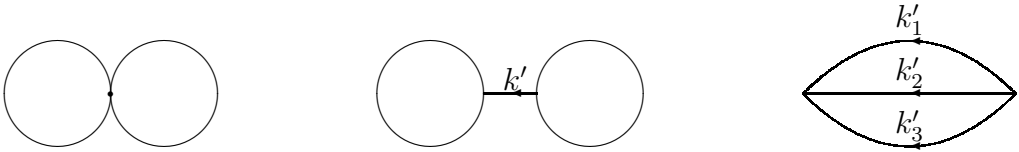

The function $\I(x)$ arises from each loop at a single vertex, or equivalently by transforming the normal ordering $::_a$ in terms of operators which create plane waves, used in the definition of the Hamiltonian, to a normal ordering in terms of operators which create normal modes $::_b$.  It solves the equation
\beq
\partial_x \I(x)=\dpin{k}\frac{1}{2\omega_k}\partial_x\left|g_{k}(x)\right|^2 \label{di}
\eeq
and tends asymptotically to 0.

We have used the identities
\beq
 V_{BBk}=-\frac{\omega_k^2}{\sqrt{Q_0}}\Delta_{kB}\hsp
 V_{Bk_1k_2}=\frac{\omega_{k_2}^2-\omega_{k_1}^2}{\sqrt{Q_0}}\Delta_{k_1 k_2} \label{vid}
 \eeq
to write (\ref{q2}) in several equivalent forms.  The expressions on the right are more complicated, but following \cite{normal} their diagrammatic interpretation, shown in Fig.~\ref{q2fig}, is more clear.  There it is explained that $Q_2^{(1)}$ corresponds to two loops at a point, $Q_2^{(2)}$ to two loops connected by an internal line, $Q_2^{(3)}$ corresponds to two points connected by 3 internal lines and the next two diagrams are obtained from the third by replacing, respectively, one or two normal mode internal lines by one or two zero-mode internal lines.  These diagrams are each UV-finite, as loops at a point lead to factors of the finite function $\I(x)$.  In the more standard diagrammatic approach of Refs.~\cite{vega,verwaest}, which use a UV cutoff, each of these diagrams corresponds to a UV-finite sum of individually divergent diagrams including diagrams with counterterms.  However, in our case no UV cutoff or counterterms are needed as normal ordering in (\ref{hd}) has already removed all UV divergences.

\section{IR-Finite Contributions} \label{finsez}

\subsection{The $\phi^4$ Double Well}

Now let us specialize to the $\phi^4$ double well theory, corresponding to the potential
\beq
V[\sqrt{\lambda}\phi(x)]=\frac{\lambda\phi^2}{4}\left(\sqrt{\lambda}\phi(x)-\b\sqrt{8}\right)^2
\eeq
and stationary classical kink solution
\beq
f(x)=\b \sqrt\frac{2}{\lambda}\left(1+\tanh(\b x)\right). \label{f}
\eeq
At the vacua $\phi=0$ and $\phi=\beta\sqrt{8/\lambda}$, we define $m^2$ to be the meson mass squared and so
\beq
m=2\b.
\eeq
The continuum normal modes, shape mode and zero mode are
\bea
g_k(x)&=&\frac{e^{-ikx}}{\ok{} \sqrt{k^2+\b^2}}\left[k^2-2\b^2+3\b^2\sech^2(\b x)-3i\b k\tanh(\b x)\right]\label{nmode}\\
g_S(x)&=&-i\sqrt\frac{3\b}{2}\tanh(\b x)\sech(\b x)\hsp
g_B(x)=\frac{\sqrt{3\b}}{2}\sech^2(\b x)\nonumber
\eea
and have frequencies
\beq
\omega_k=\sqrt{4\b^2+k^2}\hsp
\omega_S=\b\sqrt{3}\hsp
\omega_B=0.
\eeq

One easily finds the derivatives
\beq
\V3=6\sqrt{2\lambda}\b \tanh(\b x)\hsp
\V4=6\lambda
\eeq
and, solving (\ref{di}), the loop function
\beq
\I(x)=\frac{1}{4\sqrt{3}}\sech^2(\b x)-\frac{3}{8\pi}\sech^4(\b x). \label{ieq}
\eeq

\subsection{Some Useful Integrals}

In what follows, all derivatives over $x$ will be performed analytically using
\bea
\int dx e^{-ikx}\sech^{2n}(\b x)&=&\left\{
\begin{array}{cl}
2\pi\delta(k) &  {\rm{\ \ \ if}}\  n=0 \\ \frac{\pi}{(2n-1)!k}\left[\prod_{j=0}^{n-1}\left(\frac{k^2}{\b^2}+(2j)^2\right)\right]\ck   & {\rm{\ \ \ if}}\ n>0
\end{array}
\right.\nonumber\\
\int dx e^{-ikx}\sech^{2n+1}(\b x)&=& \frac{\pi}{(2n)!\b}\left[\prod_{j=0}^{n-1}\left(\frac{k^2}{\b^2}+(2j+1)^2\right)\right]\sk \nonumber\\
\int dx e^{-ikx}\sech^{2n}(\b x)\tanh(\b x)&=&-i\frac{\pi}{(2n)!\b}\left[\prod_{j=0}^{n-1}\left(\frac{k^2}{\b^2}+(2j)^2\right)\right]\ck\nonumber  \\
\int dx e^{-ikx}\sech^{2n+1}(\b x)\tanh(\b x)&=& -i \frac{\pi k}{(2n+1)!\b^2}\left[\prod_{j=0}^{n-1}\left(\frac{k^2}{\b^2}+(2j+1)^2\right)\right]\sk .
\eea

\subsection{The Last Term}

The energy of the kink ground state is expressed in (\ref{q2}) as $\sum_{i=1}^5 Q_2^{(i)}$.  The simplest term to evaluate is the fifth.  One need only use
\beq
f^{\prime\prime}(x)=-\frac{2}{\sqrt{\lambda}}\b^3\sech^2(\b x)\tanh(\b x)
\eeq
to obtain
\beq
\int dx \left|f^{\prime\prime}(x)\right|^2=\frac{4}{\lambda}\b^6\int dx \left(\sech^4(\b x)-\sech^6(\b x)\right)=\frac{16}{15}\frac{\b^5}{\lambda}.
\eeq
Alternately one may evaluate $Q_2^{(5)}$ using
\beq
\Delta_{SB}=i\pi \frac{3\b}{8\sqrt{2}}\hsp
\Delta_{kB}=i\pi \frac{\sqrt{3}}{8}\frac{k^2\ok{}}{\b^{3/2}\sqrt{\b^2+k^2}}\ck.
\eeq
The result is the same and will be summarized in Sec~\ref{numsez}.

\subsection{The Penultimate Term}

The contribution $Q_2^{(4)}$ can be found by inserting 
\bea
\Delta_{kS}&=&-i\pi \frac{\sqrt{3}}{4\sqrt{2}}\frac{(3\b^2+k^2)\sqrt{\b^2+k^2}}{\b^{3/2}\ok{}}\sk\label{delta}\\
\Delta_{k_1k_2}&=&i \pi (k_1-k_2)\delta(k_1+k_2)+i \pi \frac{3}{4}\left(\frac{\ok{1}}{\ok{2}}-\frac{\ok{2}}{\ok{1}}\right)\frac{4\b^2+k_1^2+k_2^2}{\sqrt{\b^2+k_1^2}\sqrt{\b^2+k_2^2}}\csch\left(\frac{\pi(k_1+k_2)}{2\b}\right)\nonumber
\eea
into (\ref{q2}) and integrating over continuum modes $k$ and adding the contribution from the shape mode $S$.  The contribution from the Dirac delta function vanishes, as it is multiplied by zero in (\ref{q2}).

\subsection{Contractions with the Potential}

The vertex factors are defined in (\ref{vf}).  Those involving a zero-mode are
\bea
V_{BBB}&=&V_{SSB}=0\hsp
V_{SBB}=-i\pi \frac{9\sqrt{3\lambda}}{32}\beta^{3/2}\label{vfin}\\
V_{kBB}&=&-i\pi \frac{3\sqrt{\lambda}}{16\sqrt{2}}\frac{k^2\ok{}^3}{\b^3\sqrt{\b^2+k^2}}\ck\nonumber\\
V_{kSB}&=&i\pi \frac{3\sqrt{\lambda}}{16}\frac{(3\b^2+k^2)(k^2+\b^2)^{3/2}}{\b^3\ok{}}\sk\nonumber\\
V_{k_1k_2B}&=&-i\pi \frac{3\sqrt{3\lambda}}{8\sqrt{2}}\frac{\left(\ok1^2-\ok2^2\right)^2(4\b^2+k_1^2+k_2^2)}{\b^{3/2}\ok1\ok2\sqrt{\b^2+k_1^2}\sqrt{\b^2+k_2^2}}\csch\left(\frac{\pi(k_1+k_2)}{2\b}\right).\nonumber
\eea
These are consistent with the identities (\ref{vid}).

Those involving the loop factor $\I(x)$, given in (\ref{ieq}), are
\bea
V_{\I\I}&=&\frac{\lambda}{70\b}\left(1-\frac{4\sqrt{3}}{\pi}+\frac{54}{\pi^2}\right)\label{vii}\\
V_{\I k}&=&i\frac{\sqrt{\lambda}}{32\sqrt{6}}\frac{k^2\omega_k}{\b^4\sqrt{\b^2+k^2}}\left[2\pi(-2\b^2+k^2)+3\sqrt{3}\omega_k^2\right]\ck\nonumber\\
V_{\I S}&=&i\frac{3\lambda}{64}\sqrt{\b}(3\sqrt{3}-2\pi).\nonumber
\eea
These yield $Q_2^{(1)}$ and $Q_2^{(2)}$.  So the rest of this note will be concerned with $Q_2^{(3)}$.

The only divergence arises from the term with three continuum normal modes.  The remaining finite terms are
\bea
V_{SSS}&=&i\pi\frac{9\sqrt{3\lambda}}{16}\b^{3/2}\hsp
V_{kSS}=i\pi\frac{3\sqrt{\lambda}}{8\sqrt{2}}\frac{k^2\ok{}(2\b^2-k^2)}{\b^3\sqrt{\b^2+k^2}}\ck\\
V_{k_1k_2S}&=&-i\pi \frac{3\sqrt{3\lambda}}{8}\frac{\left(17\b^4-(\ok1^2-\ok2^2)^2\right)(\b^2+k_1^2+k_2^2)+8\b^2k_1^2k_2^2}{\b^{3/2}\ok1\ok2\sqrt{\b^2+k_1^2}\sqrt{\b^2+k_2^2}}\sech\left(\frac{\pi(k_1+k_2)}{2\b}\right).\nonumber
\eea

\subsection{The Divergent Term}

Although we use (\ref{q2}) to evaluate the energy of the kink ground state, one could also use it to evaluate the vacuum ground state.  One would simply set $f(x)$ to be the corresponding expectation value of $\phi(x)$ and the $g_k(x)$ would be the corresponding linearized solutions, which are just plane waves.  Only the third term $Q_2^{(3)}$ would be nonzero and it would be divergent.  The problem is that $V_{k_1k_2k_3}$ diverges when $\sum_i k_i=0$.  In that case it would be proportional to a delta function reflecting momentum conservation in the third figure in Fig.~\ref{q2fig}.

The infrared behavior of the kink sector is identical, as at long distances the only effect of the kink is a phase-shift which is not relevant to this divergence.  The continuum normal modes tend asymptotically to plane waves, albeit shifted, and the potential tends to exactly the same constant as in the vacuum case.  Therefore, for small $\sum_i k_i$, the symbol $V_{k_1k_2k_3}$ approaches the vacuum value.  This is good news, as it means that the kink energy and the vacuum energy have the same divergence, so their difference, the kink mass, is finite.

The long distance divergence of $V_{k_1k_2k_3}$ arises from the fact that at small $\sum_i k_i$ it tends to the integral of a constant.  Thus, to regularize this divergence, we must study the integrand, which we will call $\sigma(x)$
\bea
V_{k_1k_2k_3}&=&\int dx \sigma_{k_1k_2k_3}(x)=\sum_{I=0}^3\sum_{J=0}^1 V_{k_1k_2k_3}^{IJ}\hsp
V_{k_1k_2k_3}^{IJ}=\int dx \sigma_{k_1k_2k_3}^{IJ}(x)\nonumber\\
\sigma_{k_1k_2k_3}(x)&=&\V3 g_{k_1}(x)g_{k_2}(x)g_{k_3}(x)=\sum_{I=0}^3\sum_{J=0}^1 \sigma_{k_1k_2k_3}^{IJ}(x).\label{sdef}
\eea
The indices $I$ and $J$ describe the $x$-dependence.  We separate out the $k$-dependence, the $x$-dependence and a universal $k$-dependent coefficient by introducing yet more notation
\bea
 \sigma_{k_1k_2k_3}^{IJ}(x)&=&\cc_{k_1k_2k_3}\Phi_{k_1k_2k_3}^{IJ}e^{-ix(k_1+k_2+k_3)}\sech^{2I}(\b x)\tanh^J(\b x) \label{phidef}
 \\
\cc_{k_1k_2k_3}&=&6\sqrt{2\lambda}\frac{\b}{\ok1\ok2\ok3\sqrt{\b^2+k_1^2}\sqrt{\b^2+k_2^2}\sqrt{\b^2+k_3^2}}.\nonumber
\eea
For future reference we will use a hat to define functions restricted to $\sum k_i=0$
\beq
\hp_{k_1k_2}^{IJ}=\Phi_{k_1,k_2,-k_1-k_2}^{IJ}\hsp
\hat{\cc}_{k_1k_2}=\cc_{k_1,k_2,-k_1-k_2}.
\eeq

The $x$ integrals may be performed analytically, yielding
\bea
V_{k_1k_2k_3}^{00}&=&\cc_{k_1k_2k_3}\Phi_{k_1k_2k_3}^{00}2\pi\delta(k_1+k_2+k_3)=\hat{\cc}_{k_1k_2}\hp_{k_1k_2}^{00}2\pi\delta(k_1+k_2+k_3)\\
V_{k_1k_2k_3}^{10}&=&\cc_{k_1k_2k_3}\Phi_{k_1k_2k_3}^{10}\frac{\pi\kkk}{\b^2}\csk\nonumber\\
V_{k_1k_2k_3}^{20}&=&\cc_{k_1k_2k_3}\Phi_{k_1k_2k_3}^{20}\frac{\pi\kkk}{6\b^2}\left(\frac{\left(\kkk\right)^2}{\b^2}+4\right)\csk\nonumber\\
V_{k_1k_2k_3}^{30}&=&\cc_{k_1k_2k_3}\Phi_{k_1k_2k_3}^{30}\frac{\pi\kkk}{120\b^2}\left(\frac{\left(\kkk\right)^2}{\b^2}+4\right)\left(\frac{\left(\kkk\right)^2}{\b^2}+16\right)\csk\nonumber
\eea
and
\bea
V_{k_1k_2k_3}^{01}&=&-i\cc_{k_1k_2k_3}\Phi_{k_1k_2k_3}^{01}\frac{\pi}{\b}\csk\\
V_{k_1k_2k_3}^{11}&=&-i\cc_{k_1k_2k_3}\Phi_{k_1k_2k_3}^{11}\frac{\pi\left(\kkk\right)^2}{2\b^3}\csk\nonumber\\
V_{k_1k_2k_3}^{21}&=&-i\cc_{k_1k_2k_3}\Phi_{k_1k_2k_3}^{21}\frac{\pi\left(\kkk\right)^2}{24\b^3}\left(\frac{\left(\kkk\right)^2}{\b^2}+4\right)\csk\nonumber\\
V_{k_1k_2k_3}^{31}&=&-i\cc_{k_1k_2k_3}\Phi_{k_1k_2k_3}^{31}\frac{\pi\left(\kkk\right)^2}{720\b^3}\left(\frac{\left(\kkk\right)^2}{\b^2}+4\right)\left(\frac{\left(\kkk\right)^2}{\b^2}+16\right)\csk.\nonumber
\eea
However if we simply substitute these results into (\ref{q2}), the terms containing $V^{0J}V^{0 J\p}$ would be divergent, even after integration over momenta.  One delta function or simple pole could be used to do the $k_3$ integral, but the other delta function or simple pole would leave an infinite result.  This is of course to be expected, because the kink ground state really does have a infinite energy.  We will need to subtract the vacuum energy before evaluating some $x$ integrals.

The terms with $I>0$ are not divergent.  So let us decompose $V$ into three imaginary components
\beq
V_{k_1k_2k_3}=V_{k_1k_2k_3}^{00}+V_{k_1k_2k_3}^{01}+V_{k_1k_2k_3}^{F}.
\eeq
The first contains a $\delta$ function divergence, the second a simple pole and the third is finite
\bea
V_{k_1k_2k_3}^F
&=&\cc_{k_1k_2k_3}\frac{\pi\kkk}{\b^2}\csk\\
&&\times\sum_{I=1}^3\frac{1}{(2I-1)!}\left(\Phi_{k_1k_2k_3}^{I0}-\frac{\kkk}{2I\b}i\Phi_{k_1k_2k_3}^{I1}\right)\prod_{j=1}^{I-1}\left(\frac{\left(\kkk\right)^2}{\b^2}+(2j)^2\right)\nonumber\\
\hat{V}_{k_1k_2}^F&=&V_{k_1,k_2,-k_1-k_2}^F=\hat{\cc}_{k_1k_2}\frac{2}{\b}\hp_{k_1k_2}^{10}.\nonumber
\eea

Defining the symmetrized products
\bea
S_1^n&=&k_1^n+k_2^n+k_3^n\hsp 
S_2^n=(k_1k_2)^n+(k_1k_3)^n+(k_2k_3)^n\hsp
S_3^n=(k_1k_2k_3)^n\nonumber\\
S_2^{mn}&=&k_1^mk_2^n+k_1^mk_3^n+k_2^mk_3^n+k_1^nk_2^m+k_1^nk_3^m+k_2^nk_3^m
\eea
one may use (\ref{nmode}), (\ref{sdef}) and (\ref{phidef}) to calculate the coefficients of the triple product of the continuous normal modes
\bea
\Phi_{k_1k_2k_3}^{00}&=&3i\b\left[-4\b^4S_1^1+\b^2\left(2S_2^{21}+9S_3^1\right)-S_3^1S_2^1\right]\\
\Phi_{k_1k_2k_3}^{10}&=&3i\b\left[16\b^4S_1^1+\b^2\left(-5S_2^{21}-18S_3^1\right)+S_3^1S_2^1\right]\nonumber\\
\Phi_{k_1k_2k_3}^{20}&=&9i\b^3\left[-7\b^2S_1+S_2^{21}+3S_3^1\right]\hsp \Phi_{k_1k_2k_3}^{30}=27i\b^5S_1^1\nonumber\\
\Phi_{k_1k_2k_3}^{01}&=&-8\b^6+\b^4(18S_2^1+4S_1^2)+\b^2(-2S_2^2-9S_3^1S_1^1)+S_3^2
\nonumber\\
\Phi_{k_1k_2k_3}^{11}&=&3\b^2\left[12\b^4+\b^2(-15S_2^1-4S_1^2)+(S_2^2+3S_3^1S_1^1)\right]
\nonumber\\
\Phi_{k_1k_2k_3}^{21}&=&9\b^4\left[-6\b^2+(3S_2^1+S_1^2)\right]
\hsp
\Phi_{k_1k_2k_3}^{31}=27\b^6.
\nonumber
\eea

Similarly, at $\sum_i k_i=0$ one may define the symmetrized products
\beq
\hat{S}_2=\frac{k_1^2+k_2^2+(k_1+k_2)^2}{2}\hsp
\hat{S}_3=k_1k_2(k_1+k_2)
\eeq
and write the reduced coefficients
\bea
\hp_{k_1k_2}^{00}&=&-3i\b \hat{S}_3(3\b^2+\hat{S}_2)\hsp \hp_{k_1k_2}^{10}=3i\b \hat{S}_3 \left(3\b^2+\hat{S}_2\right)\hsp
\hp_{k_1k_2}^{20}=\hp_{k_1k_2}^{30}=0\nonumber\\
\hp_{k_1k_2}^{01}&=&\hat{S}_3^2-2\b^2\hat{S}_2^2-10\b^4\hat{S}_2-8\b^6\hsp
\hp_{k_1k_2}^{11}=3\b^2(4\b^2+\hat{S}_2)(3\b^2+\hat{S}_2)\nonumber\\
\hp_{k_1k_2}^{21}&=&-9\b^4(6\b^2+\hat{S}_2)\hsp \hp_{k_1k_2}^{31}=27\b^6.
\eea

Now we have completed our decomposition of the kink ground state energy $Q_2$.  In the next section we will subtract the vacuum energy and then reassemble $Q_2$ to arrive at the kink mass.

\section{Canceling IR Divergences} \label{irsez}

\subsection{Defining the Energy Density}

Our strategy for canceling the IR divergence in the two-loop kink energy $Q_2$ is to subtract the vacuum energy density from the kink energy density before performing the spatial integration.  Our first task is thus to define the kink energy density.

Recall that our state $\vac$ is an eigenstate of the kink Hamiltonian.  The corresponding eigenvalue equation is
\beq
(H\p-Q)\vac=0\hsp Q=\sum_i Q_i.
\eeq
Expanding this equation in powers of the coupling, 
\beq
\sum_{j=0}^i \left(H_{i+2-j}-Q_{\frac{i-j}{2}+1}\right)\vac_j=0.
\eeq



The first two orders do not involve the two-loop correction $Q_2$
\beq
Q_1\vac_0=H_2\vac_0\hsp 0=H_3\vac_0+H_2\vac_1.
\eeq
It appears at the third order
\beq
Q_2\vac_0=H_4\vac_0+H_3\vac_1+(H_2-Q_1)\vac_2. \label{eom2}
\eeq
We also expand the kink Hamiltonian density order by order
\beq
H_i=\int dx \ch_i(x)\hsp
\ch_{i(>2)}(x)=\frac{1}{i!}\V{i}:\phi^i(x):_a.
\eeq

In the Schrodinger picture, the fields can be expanded in any basis of functions.  We will expand them in terms of normalized kink normal modes \cite{mekink}
\bea
\phi(x)&=&\phi_C(x)+\phi_{S}(x)+\phi_{B}(x)\nonumber\\
\phi_C(x)&=&\pin{k}\frac{1}{\sqrt{2\omega_k}}\left(b_k^\dag+b_{-k}\right)g_k(x)\nonumber\\
\phi_{S}(x)&=&\frac{1}{\sqrt{2\omega_{S}}}\left(b_{S}^\dag-b_{S}\right)g_{S}(x)\nonumber\\
\phi_{B}(x)&=&\phi_0 g_{B}(x). \label{phib}
\eea
The Hamiltonian density allows us to define the functions $\rho(x)$, as the expansion of $\ch(x)\vac$ in a Fock basis
\beq
\sum_{j=0}^i \ch_{i+2-j}(x)\vac_j=\sum_{mn} \dpink{n} \rho_i^{mn}(k_1\cdots k_n;x)\phi_0^m B_{k_1}^\dag\cdots B_{k_n}^\dag\vac_0 \label{rhodef}
\eeq
where $B_k^\dag=b_k^\dag/\sqrt{2\ok{}}.$  Integrating the $i=2$ equation over $x$ and using (\ref{eom2}) one obtains
\beq
Q_2\vac_0+Q_1\vac_2=\sum_{mn} \dpink{n} \left(\int dx \rho_2^{mn}(k_1\cdots k_n;x)\right)\phi_0^m B_{k_1}^\dag\cdots B_{k_n}^\dag\vac_0.
\eeq

Choose $\vac_{i>0}$ to be orthogonal to $\vac_0$, as one is always free to do in old-fashioned perturbation theory, then project onto $\vac_0$ to obtain
\beq
Q_2=\int dx  \rho_2^{00}(x).
\eeq
This $\rho_2^{00}$ will be our definition of the kink energy density.  As the zero-mode part of the Fock basis is not orthogonal, this choice of projection was somewhat arbitrary.  It depended on our choice of basis for the space of functions of $\phi_0$.  If another basis of functions for the $\phi_0$ were used in (\ref{rhodef}), such as a set of polynomials, the kink energy density $\rho_2^{00}(x)$ could be shifted by some linear combination of the $\rho_2^{m0}(x)$.  However, as $\vac$ is an eigenstate of $H\p$, these each integrate to zero and so the resulting kink mass would be unchanged.

\subsection{Evaluating the Energy Density}

Let us expand the kink ground state $\vac$ in this same Fock basis
\beq
\vac_i=\sum_{m,n=0}^\infty \vac_i^{mn}\hsp
\vac_i^{mn}=Q_0^{-i/2}\dpink{n}\gamma_i^{mn}(k_1\cdots k_n)\phi_0^m\Bd1\cdots\Bd n\vac_0. \label{gameq}
\eeq
Here $\gamma$ are defined to be the coefficients of this expansion.  

The divergence in $Q_2$ arises from $Q_2^{(3)}$ which in turn was derived in Ref.~\cite{me2loop} from the mixing of the free ground state $\vac_0=\vac_0^{00}$ with a kink state with three virtual normal modes $\vac_1^{03}$, given by
\beq
\gamma_1^{03}=-\frac{\sqrt{Q_0}}{6\sum_i^3 \ok{i}}V_{k_1k_2k_3}. \label{g1}
\eeq
The contribution of this state to $Q_2$ arises from the $H_3\vac_1$ term in $H\p\vac$, as seen in (\ref{eom2}).  More precisely, it arises from the first term on the right hand side of
\beq
H_3=\frac{1}{6}\int dx \V3:\phi^3(x):_a=\frac{1}{6}\int dx \V3:\phi^3(x):_b+\frac{1}{2}\int dx \V3\phi(x) \I(x) \label{h3}
\eeq
where we have changed normal ordering in terms of plane wave creation operators $::_a$ to normal ordering in terms of normal mode creation operators $::_b$ using the Wick's theorem of Ref.~\cite{mewick}.  

The corresponding two-loop energy contribution comes from $\phi^3$ acting on $\vac_1^{03}$.  This contains terms of the form (\ref{g1}) where 0, 1, 2 or 3 of the $k$ are continuum modes and the other are shape modes.  For each continuum mode, the contribution to the vacuum energy arises from $\phi_C$ which contains $b_k g_{-k}(x)$.  As $g_{-k}(x)=g_k^*(x)$, this yields a factor of $g_k^*(x)$.  For each shape mode, the contribution arises from $\phi_S$ which contains $-b_S g_S(x)$, contributing a factor of $-g_S(x)$.  However due to our convention $g_{k}(-x)=g^*_{k}(x)$, and the antisymmetry of $g_S(x)$, $g_S(x)$ is imaginary an so this contribution may be written $g_S^*(x)$.  Thus the contributions from the three factors of $\phi(x)$ are $g_{k_1}^*(x)g_{k_2}^*(x)g_{k_3}^*(x)$, multiplied by the $\V3$ in $H_3$, yielding $V_{k_1k_2k_3}^*$, irrespectively of which $k$ are continuum and discrete modes.  The three factors of $1/\sqrt{2}$ from the decomposition of $\phi_c$ in (\ref{phib}) combine with the three in the $B^\dag$ in the $\vac_1^{03}$ term of (\ref{gameq}) and the $1/6$ in (\ref{h3}), yielding a $1/48$.  In all one finds
\beq
H_3\vac_1^{03}\supset Q_2^{\rm{(3)}}\vac_0\hsp Q_2^{\rm{(3)}}= -\dpink{3}\a_{k_1k_2k_3} \left|V_{k_1k_2k_3}\right|^2\vac_0
\eeq
where
\beq
\a_{k_1k_2k_3}=\frac{1}{48\ok1\ok2\ok3\left(\ok1+\ok2+\ok3\right)}.
\eeq

The divergence arises from $k_1+k_2+k_3\sim 0$ which only occurs in the real, continuous part of the integral
\beq
Q_2^{(3)}\supset  Q_2^{(33)}=-\pink{3}\a_{k_1k_2k_3} \left|V_{k_1k_2k_3}\right|^2\vac_0
\eeq
where we defined $Q_2^{(3I)}$ to be the contribution to $Q_2^{(3)}$ in which $I$ of the three normal modes are continuous.

Now we wish to tame this IR divergence by first writing it as an integral of an energy density, defined as in (\ref{rhodef}).  Using
\beq
\ch_3(x)=\frac{1}{6} \V3:\phi^3(x):_b+\frac{1}{2} \V3\phi(x) \I(x)
\eeq
one finds that the corresponding term in (\ref{rhodef}) is
\beq
\ch_3(x)\vac_1^{03}\supset \rd(x)\vac_0\hsp \rd(x)= -\pink{3}\a_{k_1k_2k_3} V_{k_1k_2k_3}\sigma_{-k_1,-k_2,-k_3}(x). \label{rho}
\eeq
This is our definition of the energy density corresponding to IR-divergent $Q_2^{(33)}$.  Indeed, as desired, formally it satisfies
\beq
Q_2^{(33)}=\int dx \rd(x)
\eeq
although this is infinite.  The expression (\ref{rho}) for the two-loop energy density corresponding to $Q_2^{(33)}$ is a main result of this work.  The derivation can easily be modified to obtain the energy density corresponding to any term, but since the other terms are already IR-finite we will not need to do this.

We will now proceed to decompose $\rd(x)$ into pieces with various $x$-dependences
\beq
\rd(x)=\sum_{I,I\p=0}^3\sum_{J,J\p=0}^1\rd_{IJI\p J\p}(x)\hsp
\rd_{IJI\p J\p}(x)=-\pink{3}\a_{k_1k_2k_3} V^{IJ}_{k_1k_2k_3}\sigma^{I\p J\p}_{-k_1,-k_2,-k_3}(x).
\eeq
The divergences lie in the $I=I\p=0$ terms.  

We will separate the finite and infinite terms
\beq
\rd(x)=\rdiv(x)+\rfin(x)\hsp \rdiv(x)=\sum_{J,J\p=0}^1\rd_{0J0 J\p}(x). \label{rhot}
\eeq
The finite part may be integrated over $x$
\bea
\qfin&=&\int dx \rfin(x)=q_{0F}+q_{1F}+q_{FF}\label{qf}\\
q_{0F}&=& -2\pink{3}\a_{k_1k_2k_3} V_{k_1k_2k_3}^{00} V_{-k_1-k_2-k_3}^{F}=-\frac{4}{\b}\pink{2}\hat{\a}_{k_1k_2} \hat{\cc}_{k_1k_2}^2\hp^{00}_{k_1k_2}\hp^{10}_{-k_1-k_2}\nonumber\\
q_{1F}&=& -2\pink{3}\a_{k_1k_2k_3} V_{k_1k_2k_3}^{01} V_{-k_1-k_2-k_3}^{F}\nonumber\\
q_{FF}&=& -\pink{3}\a_{k_1k_2k_3} V_{k_1k_2k_3}^{F} V_{-k_1-k_2-k_3}^{F}\nonumber
\eea
where we have defined
\beq
\hat{\a}_{k_1k_2}=\a_{k_1,k_2,-k_1-k_2}.
\eeq

In all $Q_2^{(33)}$ consists of seven contributions to the kink ground state energy: $q_{0F},\ q_{1F},\ q_{FF}$ and the divergent integrals of the four $\rd_{0J0 J\p}(x)$ .

In the Introduction we claimed that this approach is equivalent to adding an IR counterterm to $\ch_4(x)$, equal and opposite to the vacuum energy density $\rv(x)$.  Let us now justify that claim.  In that case, the kink Hamiltonian would be an $x$ integral over the total kink Hamiltonian density, which includes $\ch_3(x)+\ch_4(x)$.  Then the eigenvalue equation (\ref{eom2}) for $Q_2$ would include
\beq
H\p\vac =\int dx \ch\p(x)\vac\supset \int dx\left(\ch_4(x)\vac_0+\ch_3(x)\vac_1\right)\supset \int dx\left(-\rv(x)\vac_0+\ch_3(x)\vac_1^{03}\right).
\eeq
One sees that these two terms should indeed be added before performing the $x$ integration.  First we will manipulate $\rv(x)$ to cast this subtraction in a form in which we may analytically integrate over $x$.

\subsection{The Vacuum Energy Density}

Using the third derivative of the potential
\beq
V^{(3)}[\phi_0]=6\sqrt{2\lambda}\beta
\eeq
and old-fashioned perturbation theory, one finds the one-loop vacuum energy density \cite{serone,hui}
\beq
\rv(x)=-72\lambda\b^2  \int\frac{d^3p}{(2\pi)^3}\a_{p_1p_2p_3} 2\pi\delta(p_1+p_2+p_3)=-72\lambda\b^2  \pink{2} \hat{\a}_{k_1k_2}. \label{ve}
\eeq
This is independent of $x$, but we will manipulate it to introduce a spurious $x$-dependence shortly.

The identity
\beq
\left|\Phi_{k_1k_2k_3}^{00}\right|^2+\left|\Phi_{k_1k_2k_3}^{01}\right|^2=\frac{72\lambda\b^2}{\cc_{k_1k_2k_3}^2}
\eeq
allows us to replace the 72$\lambda\b^2$ in (\ref{ve}) with a sum of two more complicated terms.  We use this to define the decomposition
\beq
\rv(x)=\rv_0(x)+\rv_1(x)\hsp
\rv_J(x)=- \pink{2} \hat{\a}_{k_1k_2}\hat{\cc}^2_{k_1k_2}\left|\hat{\Phi}_{k_1k_2}^{0J}\right|^2 \label{rhovt}
\eeq
where $J$ runs over $0$ and $1$.  Intuitively these are the decomposition in terms of the contributions from the even and odd parts of $\sigma(x)$, whose integrals respectively lead to a delta function and a simple pole.

Multiplying by the identity
\beq
1=-2i{\rm{sign}}(x)\int\frac{dk_3}{2\pi}\frac{e^{ix\kkk}}{\kkk}\label{unid}
\eeq
one finds
\beq
\rv_1(x)=2i{\rm{sign}}(x)\pink{3}\frac{e^{ix\kkk}}{\kkk}\hat{\a}_{k_1k_2}\hat{\cc}_{k_1k_2}^2\left|\hat{\Phi}_{k_1k_2}^{01}\right|^2. \label{rv1}
\eeq
We will not need the analogous formula for $\rv_0$.  

The advantage of these more complicated forms is that the vacuum energy density now has the same structure, as a matrix in $k$, as the kink energy density.  Thus we can subtract the densities at each $k$ separately.  We will see that, once the vacuum energy density has been subtracted, the energy density becomes integrable over $x$ for each $n$-tuple of $k_i$.  This means that we may perform the $x$ integration before the $k$ integration, as both are anyway finite but we are only able to perform the $x$ integration analytically.

Nonetheless this trick is somewhat expensive numerically, as it yields our only three-dimensional integration over $k_i$ which does not decrease exponentially in $\kkk$.  To resolve this problem, we use the sine integral function
\beq 
{\rm{Si}}(a)=\int_0^a dt \frac{\sin(t)}{t}. 
\eeq
Using
\beq
i\left(\frac{{\rm{sign}}(x)}{2}-\frac{{\rm{Si}}(a x)}{\pi}\right)=\int_{-\infty}^{-k_1-k_2-a} \frac{dk_3}{2\pi}  \frac{e^{ix(\kkk)}}{\kkk}+\int_{-k_1-k_2+a}^{\infty} \frac{dk_3}{2\pi}  \frac{e^{ix(\kkk)}}{\kkk}
\eeq
one can replace (\ref{unid}) with
\beq
1=-2i{\rm{sign}}(x)\int_{-k_1-k_2-a}^{-k_1-k_2+a}\frac{dk_3}{2\pi}\frac{e^{ix\kkk}}{\kkk}
+\left(1-\frac{2{\rm{sign}}(x){\rm{Si}}(a x)}{\pi}\right).
\eeq
This allows one to reduce the $k_3$ range of integration
\beq
\rv_1(x)=2i{\rm{sign}}(x)\pink{2}\int_{-k_1-k_2-a}^{-k_1-k_2+a} \frac{dk_3}{2\pi} \frac{e^{ix\kkk}}{\kkk}\hat{\a}_{k_1k_2}\hat{\cc}_{k_1k_2}^2\left|\hat{\Phi}_{k_1k_2}^{01}\right|^2+R(x,a) \label{rv1r}
\eeq
where $a$ is any positive number and the remainder is
\beq
R(x,a)=- \left(1-\frac{2{\rm{sign}}(x){\rm{Si}}(a x)}{\pi}\right)\pink{2} \hat{\a}_{k_1k_2}\hat{\cc}^2_{k_1k_2}\left|\hat{\Phi}_{k_1k_2}^{0J}\right|^2.
\eeq
The remainder is easily integrated over $x$
\beq
\int dx R(x,a)=-\frac{4}{\pi a}\pink{2} \hat{\a}_{k_1k_2}\hat{\cc}^2_{k_1k_2}\left|\hat{\Phi}_{k_1k_2}^{01}\right|^2. \label{intr}
\eeq
Notice that in the limit in which $a\rightarrow 0$, this integral is infinite, as it is equal to $\int dx\rv_1(x).$


\subsection{Canceling IR Divergences}

We are now ready to subtract the vacuum energy density from the kink energy density components defined in (\ref{rhot}).  We will do this one term at a time.

\subsubsection{$\rd_{0000}(x)$}

The first contribution to the kink energy density, resulting from two even $\sigma(x)$ terms,  is
\bea
\rd_{0000}(x)&=&-\pink{3}\a_{k_1k_2k_3} V^{00}_{k_1k_2k_3}\sigma^{00}_{-k_1,-k_2,-k_3}(x)\\
&=&-\pink{3}\a_{k_1k_2k_3}\hat{\cc}_{k_1k_2}\hp_{k_1k_2}^{00}2\pi\delta(k_1+k_2+k_3)
\cc_{k_1k_2k_3}\Phi_{-k_1-k_2-k_3}^{00}e^{ix(k_1+k_2+k_3)}\nonumber\\
&=&-\pink{2}\hat{\a}_{k_1k_2}\hat{\cc}^2_{k_1k_2}\left|\hp_{k_1k_2}^{00}\right|^2=
\rv_0(x).
\nonumber
\eea

Subtracting the corresponding contribution to the vacuum energy (\ref{rhovt}) one obtains the corresponding contribution to the two-loop kink ground state mass
\beq
\rdif_{00}(x)=\rd_{0000}(x)-\rv_0(x)=0.
\eeq

\subsubsection{$\rd_{0101}(x)$}

The next contribution arises from the two odd $\sigma(x)$ terms
\bea
\rd_{0101}(x)&=&-\pink{3}\a_{k_1k_2k_3} V^{01}_{k_1k_2k_3}\sigma^{01}_{-k_1,-k_2,-k_3}(x)\\
&=&2i\pink{3}\a_{k_1k_2k_3}
\cc_{k_1k_2k_3}^2\left|\Phi_{k_1k_2k_3}^{01}\right|^2\frac{\pi}{2\b}\csk
e^{ix(k_1+k_2+k_3)}\tanh(\b x).\nonumber
\eea

Subtracting the vacuum energy density (\ref{rv1}) one finds
\bea
\rd_{0101}(x)-\rv_1(x)&=&2i\pink{3}e^{ix(k_1+k_2+k_3)}\\
&\times&\left[\a_{k_1k_2k_3}\cc_{k_1k_2k_3}^2\left|\Phi_{k_1k_2k_3}^{01}\right|^2\frac{\pi}{2\b}\csk
\tanh(\b x)\right.\nonumber\\
&&\left.
-\frac{1}{\kkk}\hat{\a}_{k_1k_2}\hat{\cc}_{k_1k_2}^2\left|\hat{\Phi}_{k_1k_2}^{01}\right|^2{\rm{sign}}(x)
\right].\nonumber
\eea
In the limit $|x|\rightarrow\infty$ the term in square brackets tends to a finite value for all $k$, as the residues of the simple poles are equal and opposite.  Therefore we may integrate over $x$ to obtain the corresponding contribution to the kink mass
\bea
q_{11}&=&\int dx\left(\rd_{0101}(x)-\rv_1(x)\right)\\
&=&4\pink{3}\left[-\a_{k_1k_2k_3}\cc_{k_1k_2k_3}^2\left|\Phi_{k_1k_2k_3}^{01}\right|^2\left(\frac{\pi}{2\b}\right)^2\csks
\right.\nonumber\\
&&\left.
+\frac{1}{\left(\kkk\right)^2}\hat{\a}_{k_1k_2}\hat{\cc}_{k_1k_2}^2\left|\hat{\Phi}_{k_1k_2}^{01}\right|^2
\right].\nonumber
\eea
The term in brackets has only a first order pole at $k_1+k_2+k_3=0$ as the residues of the second order poles cancel.   We define this integral to be the principal value, which is finite and real.  Any other prescription would lead to a finite contribution at arbitrarily small $k$, which is unphysical as long wavelength modes have measure zero support on the kink.

\subsubsection{$\rd_{0001}(x)+\rd_{0100}(x)$}

Now we have subtracted the entire vacuum energy, but we still have two divergent terms left in the kink energy.  These are the cross terms arising from one even and one odd $\sigma(x)$.  As a consistency check on our calculation, their sum must be finite.

The two terms are
\bea
\rd_{0100}(x)&=&-\pink{3}\a_{k_1k_2k_3} V^{01}_{k_1k_2k_3}\sigma^{00}_{-k_1,-k_2,-k_3}(x)\\
&=&2i\pink{3}\a_{k_1k_2k_3}
\cc_{k_1k_2k_3}^2\Phi_{k_1k_2k_3}^{01}\Phi_{-k_1-k_2-k_3}^{00}\frac{\pi}{2\b}\csk
e^{ix\kkk}\nonumber
\eea
and
\bea
\rd_{0001}(x)&=&-\pink{3}\a_{k_1k_2k_3} V^{00}_{k_1k_2k_3}\sigma^{01}_{-k_1,-k_2,-k_3}(x)\\
&=&-\pink{3}\a_{k_1k_2k_3}\hat{\cc}_{k_1k_2}\hp_{k_1k_2}^{00}2\pi\delta(k_1+k_2+k_3)
\cc_{k_1k_2k_3}\Phi_{-k_1-k_2-k_3}^{01}\tanh(\b x)\nonumber\\
&=&-\pink{2}\hat{\a}_{k_1k_2}\hat{\cc}^2_{k_1k_2}\hp_{k_1k_2}^{00}\hp_{-k_1-k_2}^{01}\tanh(\b x)
\nonumber\\
&=&-\pink{3}\hat{\a}_{k_1k_2}\hat{\cc}^2_{k_1k_2}\hp_{k_1k_2}^{00}\hp_{-k_1-k_2}^{01}\left[-i\frac{\pi}{\b}e^{ix\kkk}\csk\right]\nonumber\\
&=&-2i\pink{3}\hat{\a}_{k_1k_2}\hat{\cc}^2_{k_1k_2}\hp_{-k_1-k_2}^{00}\hp_{k_1k_2}^{01}\frac{\pi}{2\b}\csk e^{ix\kkk}.\nonumber
\eea

Their sum is
\bea
\rd_{0100}(x)+\rd_{0001}(x)&=&2i\pink{3}\frac{\pi\kkk}{2\b}\csk
e^{ix\kkk}T_{k_1k_2k_3}\\
T_{k_1k_2k_3}&=&
\frac{\a_{k_1k_2k_3}
\cc_{k_1k_2k_3}^2\Phi_{k_1k_2k_3}^{01}\Phi_{-k_1-k_2-k_3}^{00}
-\hat{\a}_{k_1k_2}\hat{\cc}^2_{k_1k_2}\hp_{k_1k_2}^{01}\hp_{-k_1-k_2}^{00}}{\kkk}.
\nonumber
\eea
The numerator vanishes linearly as $\kkk$ tends to zero, and so
\beq
\hat{T}_{k_1k_2}=T_{k_1,k_2,-k_1-k_2}=\left.\frac{\partial}{\partial{k_3}}\left(\a_{k_1k_2k_3}
\cc_{k_1k_2k_3}^2\Phi_{k_1k_2k_3}^{01}\Phi_{-k_1-k_2-k_3}^{00}\right)\right|_{k_3=-k_1-k_2}
\eeq
is finite, as is
the integrand.  Now we may integrate over $x$
\bea
q_{10}&=&\int dx\left(\rd_{0100}(x)+\rd_{0001}(x)\right)\\
&=&2i\pink{3}\frac{\pi\kkk}{2\b}\csk
T_{k_1k_2k_3}2\pi\delta\left(\kkk\right)\nonumber\\
&=&2i\pink{2}\hat{T}_{k_1k_2}.\nonumber
\eea

In all, we have found five contributions to the IR finite $Q_2^{(33)}=\int dx \rd(x)$, obtained by subtracting $\rv(x)$ from the integrand at each $x$
\beq
\int dx \left(\rd(x)-\rv(x)\right)=q_{0F}+q_{1F}+q_{FF}+q_{11}+q_{10} \label{cinque}
\eeq 
each of which is finite.

\section{Numerical Integration} \label{numsez}


Now that all $x$ integration has been done analytically, we will perform the $k$ integration in (\ref{q2}) numerically.  Uncertainties will be reported in parentheses for those integrals that dominate the error budget.

Two summands require no integration over the momenta $k$ and so are easily determined analytically using (\ref{vii}) and (\ref{f}) respectively
\beq
Q_2^{(1)}=\frac{1}{560}\left(1-\frac{4\sqrt{3}}{\pi}+\frac{54}{\pi^2}\right)\frac{\lambda}{\beta}
\hsp
Q_2^{(5)}=-\frac{1}{8}\frac{9\lambda^2}{64\b^6}\frac{16}{15}\frac{\b^5}{\lambda}=-\frac{3}{80}\frac{\lambda}{\beta}.
\eeq
Numerically, these are
\beq
Q_2^{(1)}\sim 0.00761791\frac{\lambda}{\beta} \hsp
Q_2^{(5)}=-0.0375\frac{\lambda}{\beta} .
\eeq 

Next, again using (\ref{vii}), we find the second contribution
\bea
Q_2^{(2)}&=&-\frac{1}{8}\pin{k}\frac{\left|V_{\I  k}\right|^2}{\okp{}^2}-\frac{\left|V_{\I S}\right|^2}{8\omega_S^2}\\
&=&-\frac{1}{8}\frac{\lambda}{6144\b^8}\pin{k}\frac{k^4}{\b^2+k^2}\left[2\pi(-2\b^2+k^2)+3\sqrt{3}\omega_k^2\right]^2\csch^2\left(\frac{\pi k}{2\b}\right)\nonumber\\
&&-\frac{1}{8}\frac{3}{4096}\left[-2\pi+3\sqrt{3}\right]^2\frac{\lambda}{\b}\sim
(-0.000961713-0.000108182)\frac{\lambda}{\b}\nonumber\\
&\sim&-0.0010699\frac{\lambda}{\b}.\nonumber
\eea

The fourth term can be found using (\ref{delta})
\bea
Q_2^{(4)}&=&\frac{1}{16Q_0}\pink{2}\frac{\left|\left(\omega_{k_1}-\omega_{k_2}\right)\Delta_{k_1 k_2}\right|^2}{\omega_{k_1}\omega_{k_2}}+\frac{1}{8Q_0}\pink{1}\frac{\left|\left(\omega_{k_1}-\omega_{S}\right)\Delta_{k_1 S}\right|^2}{\omega_{k_1}\omega_{S}}\\
&=&\frac{27\pi^2\lambda}{2048\b^3}\pink{2}\frac{\left(\omega_{k_1}-\omega_{k_2}\right)^2\left(k_1^2-k_2^2\right)^2}{\ok1^3\ok2^3}\frac{\left(4\b^2+k_1^2+k_2^2\right)^2}{(\b^2+k_1^2)(\b^2+k_2^2)}\csch^2\left(\frac{\pi(k_1+k_2)}{2\b}\right)\nonumber\\
&&+\frac{3\sqrt{3}\pi^2\lambda}{2048\b^3}\pin{k}\frac{\left(\omega_{k}-\b\sqrt{3}\right)^2}{\b\omega_{k}}
\frac{(3\b^2+k^2)^2(\b^2+k^2)}{\b^{3}\ok{}^2}\sks\nonumber\\
&\sim&(0.001481577+0.002358405)\frac{\lambda}{\b}\sim 0.00383998\frac{\lambda}{\b}.\nonumber 
\eea

It remains to evaluate the third term $Q_2^{(3)}$.  Recall that this term results from two vertices with three normal modes each.  The normal modes may be the shape mode $S$ or continuum modes.  Let us first decompose the result into summands involving various numbers of continuum modes
\bea
Q_2^{(3)}&=&\sum_{k=0}^3 Q_2^{(3k)}\hsp
Q_2^{(30)}=-\frac{1}{48} \frac{\left|V_{SSS}\right|^2}{3\omega_{S}^4}\\
Q_2^{(31)}&&=-\frac{1}{16}\pink{1} \frac{\left|V_{SSk_1}\right|^2}{\omega_{k_1}\omega_{S}^2\left(\omega_{k_1}+2\omega_S\right)}\nonumber\\
Q_2^{(32)}&&=-\frac{1}{16}\pink{2} \frac{\left|V_{Sk_1k_2}\right|^2}{\omega_{k_1}\ok2\omega_{S}\left(\omega_{k_1}+\ok2+\omega_S\right)}\nonumber\\
Q_2^{(33)}&&=q_{0F}+q_{1F}+q_{FF}+q_{11}+q_{10}.\nonumber
\eea
Only the term with three continuum modes is divergent.  The others can be found using (\ref{vfin})
\bea
Q_2^{(30)}&=&-\frac{3\pi^2}{4096} \frac{\lambda}{\b}\sim -0.00722871  \frac{\lambda}{\b}\\
Q_2^{(31)}&=&-\frac{3\pi^2\lambda}{2048\b^8}\pin{k} \frac{k^4 \ok{}(2\b^2-k^2)^2}{\left(\omega_{k}+2\sqrt{3}\b\right)(\b^2+k^2)}\cks\sim -0.0008745152 \frac{\lambda}{\b}  \nonumber\\
Q_2^{(32)}&=&-\frac{9\sqrt{3}\pi^2\lambda}{1024\b^4}\pink{2} \frac{\left[ \left(17\b^4-(k_1^2-k_2^2)^2\right)(\b^2+k_1^2+k_2^2)+8\b^2k_1^2k_2^2\right]^2\sech^2\left(\frac{\pi(k_1+k_2)}{2\b}\right)}{\ok1^3\ok2^3\left(\omega_{k_1}+\ok2+\sqrt{3}\b\right)(\b^2+k_1^2)(\b^2+k_2^2)}\nonumber\\
&\sim&-0.0311512(1) \frac{\lambda}{\b}.\nonumber
\eea

Finally we are ready for the five terms (\ref{cinque}) involving three continuum modes.  These are the only terms which have analogs in the vacuum sector, defined by replacing each $g_k$ with the corresponding plane wave and the kink Hamiltonian with the defining Hamiltonian, so that all $\Delta$ vanish and $V$ is proportional to a Dirac delta function.   Thus these are the only terms for which we subtract the vacuum contribution.  As an abuse of notation, we continue to call this contribution $Q_2^{(33)}$ after this infinite subtraction.

A single divergent term in $V$ yields a delta function or simple pole which can be used to do the $k_3$ integral.  However $Q_2^{(33)}$ is quadratic in $V$ and so it is the second divergence in $V$ which causes an infinity in $Q_2^{(33)}$.  This means that the three terms (\ref{qf}) which involve zero or one divergent terms $V^{00}$ or $V^{01}$ in the two $V$ factors, and finite terms $V^F$ from the others, are manifestly finite and require no vacuum subtraction.  Let us begin with these three terms.

The first contribution arises from the cross term between one interaction $V^{00}$ proportional to Dirac delta function in momentum space, corresponding to the position-independent part of the triple product of three normal modes, with the finite terms $V^F$ in the triple product.  As the delta function can be used to do the $k_3$ integral, this contribution is given by a two-dimensional integral
\bea
q_{0F}
&=&54\b^3\lambda\pink{2}\frac{k_1^2k_2^2(k_1+k_2)^2(3\b^2+k_1^2+k_1k_2+k_2^2)^2}{\ok1^3\ok2^3\omega^3_{k_1+k_2}\left(\ok1+\ok2+\omega_{k_1+k_2}\right)(\b^2+k_1^2)(\b^2+k_2^2)(\b^2+(k_1+k_2)^2)}
\nonumber\\
&\sim&0.0375390(1) \frac{\lambda}{\b}.
\eea

For the three-dimension terms that follow, we will often encounter the combination
\beq
\a_{k_1k_2k_3}\cc^2_{k_1k_2k_3}= \frac{3\lambda\b^2}{2\ok1^3\ok2^3\ok3^3\left(\ok1+\ok2+\ok3\right)(\b^2+k_1^2)(\b^2+k_2^2)(\b^2+k_3^2)}.
\eeq
The next finite term is the cross term between the antisymmetric $V^{01}$ and the finite $V^F$
\bea
q_{1F}
&=&\frac{3\pi^2\lambda}{\b}\pink{3}\frac{\Phi_{k_1k_2k_3}^{01}\left(\kkk\right)\csks}{\ok1^3\ok2^3\ok3^3\left(\ok1+\ok2+\ok3\right)(\b^2+k_1^2)(\b^2+k_2^2)(\b^2+k_3^2)} \\
&&\times\sum_{I=1}^3\frac{1}{(2I-1)!}\left(-i\Phi_{k_1k_2k_3}^{I0}-\frac{\kkk}{2I\b}\Phi_{k_1k_2k_3}^{I1}\right)\prod_{j=1}^{I-1}\left(\frac{\left(\kkk\right)^2}{\b^2}+(2j)^2\right)
\nonumber\\
&\sim&0.006070(1) \frac{\lambda}{\b}.
\eea
Here the integral is evaluated according to the principal value prescription, as it must be real and no finite contribution from spatial infinity, corresponding to $\kkk=0$, is expected.

The last finite term corresponds to the product of the two finite terms $V^F$
\bea
q_{FF}
&=& -\frac{3\pi^2\lambda}{2\b^2}\pink{3}\frac{(\kkk)^2\csks}{\ok1^3\ok2^3\ok3^3\left(\ok1+\ok2+\ok3\right)(\b^2+k_1^2)(\b^2+k_2^2)(\b^2+k_3^2)}\nonumber\\
&&\times\left[\sum_{I=1}^3\frac{1}{(2I-1)!}\left(i\Phi_{k_1k_2k_3}^{I0}+\frac{\kkk}{2I\b}\Phi_{k_1k_2k_3}^{I1}\right)\prod_{j=1}^{I-1}\left(\frac{\left(\kkk\right)^2}{\b^2}+(2j)^2\right)\right]^2\nonumber\\
&\sim&-0.0163976(4) \frac{\lambda}{\b}.
\eea

Recall that, only in the case of $q_{11}$, the integral decreases as $1/k$ in all three directions and so the integral is truly three-dimensional.  In all other three-dimensional integrals, and in the kink sector contribution to $q_{11}$, the $k_3$ integrand converges exponentially as the result of a csch function.  However, as explained above, the $k_3$ integration of the vacuum sector contribution can be performed analytically above any given cutoff $a$, leaving the two-dimensional integral given in Eq.~(\ref{intr}).  We therefore choose this cutoff to be at least $\kkk=12$, so that the kink energy term, as it is exponentially suppressed in $\kkk$, is easily calculated to within our error budget.   Above this threshold the choice of cutoff has negligible effect on our result
\bea
q_{11}&=&6\lambda \b^2\pink{3}\frac{1}{\ok1^3\ok2^3(\b^2+k_1^2)(\b^2+k_2^2)}\left[-\frac{\left|\Phi_{k_1k_2k_3}^{01}\right|^2\left(\frac{\pi}{2\b}\right)^2\csks}{\ok3^3\left(\sum_{j=1}^3\ok{j}\right)(\b^2+k_3^2)}
\right.\nonumber\\
&&\left.
+\frac{1}{\left(\kkk\right)^2}\frac{\left|\hat{\Phi}_{k_1k_2}^{01}\right|^2}{\omega_{k_1+k_2}^3\left(\ok1+\ok2+\omega_{k_1+k_2}\right)(\b^2+(k_1+k_2)^2)}
\right]\nonumber\\
&\sim&6\lambda \b^2\pink{2}\int_{-k_1-k_2-a}^{-k_1-k_2+a}\frac{dk_3}{2\pi}\frac{1}{\ok1^3\ok2^3(\b^2+k_1^2)(\b^2+k_2^2)}\nonumber\\
&&\times\left[-\frac{\left|\Phi_{k_1k_2k_3}^{01}\right|^2\left(\frac{\pi}{2\b}\right)^2\csks}{\ok3^3\left(\sum_{j=1}^3\ok{j}\right)(\b^2+k_3^2)}
\right.\nonumber\\
&&\left.
+\frac{1}{\left(\kkk\right)^2}\frac{\left|\hat{\Phi}_{k_1k_2}^{01}\right|^2}{\omega_{k_1+k_2}^3\left(\ok1+\ok2+\omega_{k_1+k_2}\right)(\b^2+(k_1+k_2)^2)}
\right]\nonumber\\
&&+\frac{4}{\pi a}\pink{2} \hat{\a}_{k_1k_2}\hat{\cc}^2_{k_1k_2}\left|\hat{\Phi}_{k_1k_2}^{01}\right|^2\sim 0.033043(2) \frac{\lambda}{\b}.
\eea

Finally the cross term between the two divergent $V$ yields
\bea
q_{10}&=&-3\lambda\b^2\pink{2}\frac{\partial}{\partial{k_3}}\left.\left( \frac{i\Phi_{k_1k_2k_3}^{00}\Phi_{k_1k_2k_3}^{01}}{\ok1^3\ok2^3\ok3^3\left(\ok1+\ok2+\ok3\right)(\b^2+k_1^2)(\b^2+k_2^2)(\b^2+k_3^2)}\right)\right|_{k_3=-k_1-k_2}\nonumber\\
&\sim&0.0124290(1)\frac{\lambda}{\b}.
\eea

\section{Concluding Remarks} \label{compsez}

Adding all 12 contributions one finds that the two-loop correction to the $\phi^4$ kink mass is
\beq
Q_2\sim 0.006316(2)\frac{\lambda}{\b}=0.012633(5)\frac{\lambda}{m}.
\eeq
This is positive, in agreement with small $\lambda$ lattice results \cite{lat98}, instantaneous frame Fock space truncation results \cite{slava,bajnok} and Borel resummation \cite{serone}.  It also agrees with the preferred value of the mass correction in light front \cite{vary03} and conformal space \cite{mussardo} truncations.  On the other hand, the lattice study in Ref.~\cite{latt09} found a kink mass beneath the one-loop result.  Fig.~8 of Ref.~\cite{slava} appears to show a mass correction of $0.018\pm 0.008$ at $\lambda=0.4$ and $m=1$, if the dot size is to be interpreted as the error bar, which may be compared with our result of $Q_2\sim 0.005$.  Our result also lies just beyond the lower error bar of Fig.~11 of Ref.~\cite{serone}.  In both cases, this light tension is smaller than some of the individual mass contributions, such as $Q_2^{(5)}$ which,  at $\lambda=0.4$ and $m=1$, contributes $-0.03$ to $Q_2$.

In principle, the methods cited above include nonperturbative and in the case of Ref.~\cite{serone} also higher order perturbative information, and so their results are more reliable as the coupling grows.  In practice, the uncertainties in all of these studies are of the same order as the difference between the calculated kink mass and the one loop result (\ref{q1}).  In contrast, the uncertainty on our $Q_2$ is several thousand times smaller than its central value.  In that sense, our method offers far more precise results, although as the coupling increases the accuracy will be compromised by higher order and also nonperturbative corrections, with the latter arising from virtual kink-antikink pair creation.

While $Q_2$ is positive, the coefficient is not large enough to invalidate the observation of Ref.~\cite{slava} that a naive extrapolation of the mass formula, together with the semiclassical calculation of the bound state masses in Ref.~\cite{mussardo}, suggests that at large enough coupling, all bound states of kinks and so potentially all topologically trivial particles are no longer in the spectrum.  Similarly, the coupling at which the meson mass is twice the kink mass, and so the meson may decay \cite{mussardo06,bajnok}, is hardly affected.  Of course there is no reason to trust our perturbative analysis beyond weak coupling.

Chang duality \cite{chang} implies that the vacuum Hamiltonian, at each value of $\lambda/\b^2$ beneath about 8, is equal to the vacuum Hamiltonian at a large value of $\lambda/\b^2$, as the shift in the classical Hamiltonian is exactly compensated by the change in normal ordering as the mass changes.   In other words, each quantum Hamiltonian arises from two distinct classical Hamiltonians with $\b^2>0$.   We do not expect our semiclassical expansion to be reliable at such large couplings, although the positive $Q_2$ found here is consistent with the possibility that the kink belongs to a Chang-symmetric family of Hamiltonian eigenstates which exists at all couplings and is continuous in the coupling.  This possibility is also weakly supported by the fact that that the kink mass appears to become flat at the right of Fig.~11 of Ref.~\cite{serone}.   Stronger evidence arises from the $\b^2<0$ Chang dual, whose Hamiltonian we recall is the same operator.  Ref.~\cite{serone} found that its mass gap is well-defined throughout this region and coincides with the kink mass when the $\b^2>0$ theory is at coupling below the critical coupling.   In an intermediate regime of couplings including the self-dual point, one expects that the $\Z_2$ symmetry, whose breaking is responsible for the classical kink solution, is restored.  Thus any continuation of the kink state in that region would be quite interesting.

Moreover, this duality implies that our perturbative kink states, found at weak coupling, are also eigenstates of a strongly coupled Hamiltonian to the same precision.  In particular, they become exact eigenstates in the infinite $\lambda/\b^2$ limit.  In this limit they do not become the semiclassical solitons of the strongly coupled theory, which one expects to be deformed beyond recognition by quantum corrections.  For example, the $\df$ in the construction uses the classical kink solution $f(x)$ of the original classical Hamiltonian, which is not a solution of the classical equations of motion of the dual Hamiltonian.   These may therefore provide an example of a quantum soliton unrelated to any classical solution of the same system.  As this a property that any monopoles appearing in Yang-Mills would also possess, they could prove to be a useful testing ground.


\section* {Acknowledgement}

\noindent
JE is supported by the CAS Key Research Program of Frontier Sciences grant QYZDY-SSW-SLH006 and the NSFC MianShang grants 11875296 and 11675223.   JE also thanks the Recruitment Program of High-end Foreign Experts for support.

\end{document}